\documentclass[aps,pra,twocolumn,amsmath,amssymb,longbibliography]{revtex4-1}
\usepackage{amssymb}
\usepackage{amsmath}
\usepackage{bm}
\usepackage{hyperref}
\usepackage{graphicx}
\usepackage{subfigure}
\usepackage{color}

\newcommand{\ket}[1]{|#1\rangle}
\newcommand{\bra}[1]{\langle #1|}
\newcommand{\inp}[2]{\langle #1|#2\rangle}

\begin{document}

\title{Optimal joint estimation of multiple Rabi frequencies}


\author{Hongzhen Chen}
\author{Haidong Yuan}
\email{hdyuan@mae.cuhk.edu.hk}
\affiliation{Department of Mechanical and Automation Engineering,\\The Chinese University of Hong Kong, Shatin, Hong Kong}


\date{\today}

\begin{abstract}
We study the joint estimation of multiple Rabi frequencies in a multi-level system. By analytically identifying the optimal probe state and the optimal measurement, we obtain the highest precision limit for the joint estimation of multiple Rabi frequencies. We show that the joint estimation does not always outperform the separate estimation. There exist different regimes where the joint estimation can either outperform or underperform the separate estimation. We further show that if additional adaptive quantum controls are allowed then the advantage of the joint estimation over the separate estimation can be reestablished.
\end{abstract}
\pacs{}

\maketitle

\section{Introduction}
Quantum metrology, which makes use of quantum mechanical effects to achieve higher precision, is gaining increasing attention for its broad applications in many areas \cite{Degen2017,Braun2018}, such as atomic clocks \cite{Roos:2006dl,Leroux:2010jg,LouchetChauvet:2010fs,Hosten2016,Pezze2018}, spectroscopy \cite{Leibfried:2004ii,Schmidt:2005jq,Davis2016,Shaniv2018}, magnetometry \cite{Koschorreck:2010gj,Sewell:2012cn,Brask2015,Schmitt2017}, gravitational-wave detection \cite{Schnabel:2010cr,Aasi2013,Adhikari2014,Abbott2016}, etc.
Much progress has been made in the single parameter quantum estimation \cite{Giovannetti:2004jg,Giovannetti:2006cr,Giovannetti:2011jk}.
However, in many practical applications such as microscopy and imaging \cite{Preza:1999ij,Brida2010,Low2015,Lupo2016,Nair2016,McConnell2017,Nair2018}, there are typically more than one parameter.
This puts a high demand on better understandings of multi-parameter quantum estimation.
There have been recent studies on joint estimation of multiple parameters in various specific settings, such as multiple phase estimation under commuting dynamics \cite{Humphreys:2013gz}, estimation of  multidimensional fields \cite{Baumgratz:2016ii}, estimation of multiple parameters in quantum states \cite{Gill:2000cd,Li:2016gc} and unitary operators \cite{Acin2001,Fujiwara:2001dk,Ballester:2004kr,Bartlett2007,Liu2015,Zhou:2015gf,Berry:2015bb}, joint estimatin of the phase and decoherence rate \cite{Vidrighin:2014fa,Crowley:2014ja}, as well as joint estimation of two decoherence rates \cite{Monras:2011ip}.
However, the optimal performance for multi-parameter quantum estimation is still not well understood\cite{Chen2017,Pezze:2017ez,2018arXiv180207587Y}.

A distinct feature of multi-parameter quantum estimation is the tradeoff among the precisions of different parameters \cite{Vidrighin:2014fa,Gill:2000cd,Li:2016gc,Zhu2018, Hou2018,Crowley:2014ja,Baumgratz:2016ii,Pezze:2017ez}, as the optimal probe states and the optimal measurements for different parameters are typically different. Such trade-offs are usually quantified by a particular figure of merit, typically a combination of the weighted precisions of different parameters, one can then optimize the probe state and the measurement according to the chosen figure of merit. This turns out to be a hard problem.
While general methods on the identification of the optimal probe states exist for the single-parameter quantum estimation\cite{Yuan:2017fr}, the probe states for multi-parameter quantum estimation are typically chosen heuristically\cite{Ballester:2004kr,Monras:2011ip,Baumgratz:2016ii}, the optimal probe states are only identified for very few cases\cite{Humphreys:2013gz,Crowley:2014ja,Yuan:2016hf}.

In this article, we study the joint estimation of multiple Rabi frequencies. We start with the joint estimation of two Rabi frequencies in a three-level system. By explicitly optimizing the probe state and the measurement, we analytically obtain the highest precision for the joint estimation of two Rabi frequencies. We find that, in contrast to the expectation, the joint estimation does not always outperform the separate estimation. There exist different regimes that the joint estimation can either outperform or underperform the separate estimation. This enriches the understandings on the relationship between the joint and the separate estimation. We then consider adding optimal quantum controls in the scheme and show that with the adaptive quantum controls, the joint estimation can restore the advantage over the separate estimation.

The article is organized as follows: in Sec.\ref{sec:multi} we make a brief introduction of the basic tools for the multi-parameter quantum estimation; in Sec.\ref{sec:QCRB} we derive the precision limits for the joint estimation of two Rabi frequencies in a three-level system by explicitly obtaining the optimal probe state and the optimal measurement scheme; we then compare the joint estimation with the separate estimation in Sec.\ref{sec:comparison}; in Sec.\ref{sec:control} we show that with optimal adaptive controls the joint estimation can always outperform the separate estimation. This is then generalized to systems with more levels in Sec.\ref{sec:multilevel}; Sec.\ref{sec:conclude} concludes the paper.

\section{Multi-parameter Quantum Estimation}\label{sec:multi}

To estimate a set of parameters, $\bm{\varphi}=(\varphi_1,...,\varphi_p)$, encoded in a quantum channel $\Lambda_{\bm{\varphi}}$, one can prepare a probe state, $\rho$, and get the output state, $\rho_{\bm{\varphi}}=\Lambda_{\bm{\varphi}}(\rho)$, which contains the unknown parameters.
With a set of Positive Operator Valued Measurements (POVMs), $\{M_x\}_{x\in\Omega}$, the information of the parameters can then be extracted. From the probability distribution of the measurement results, $p_{\bm{\varphi}}(x)=Tr(\rho_{\bm{\varphi}}M_x)$, one can construct an estimator, $\bm{\hat{\varphi}}(x)=(\hat{\varphi}_1,...,\hat{\varphi}_p)(x)$.
For locally unbiased estimator with $\left<\bm{\hat{\varphi}}\right>=\sum_{x\in\Omega} \bm{\hat{\varphi}}(x) p_{\bm{\varphi}}(x)=\bm{\varphi}$ and $d\left<\bm{\hat{\varphi}}\right>/d\varphi_i=1$, $\forall i$, the precision is bounded below by the Fisher information matrix \cite{Fisher:2008gt,Cramer1946,Rao1945,Kay:1993:FSS:151045} as
\begin{equation}
  Cov(\bm{\hat{\varphi}})\ge \mathcal{I}(\bm{\varphi})^{-1},
\end{equation}
where $Cov(\bm{\hat{\varphi}})$ is the covariance matrix with its $ij$-th entry as $[Cov(\bm{\hat{\varphi}})]_{ij}=\left<(\hat{\varphi_i}-\varphi_i)(\hat{\varphi_j}-\varphi_j)\right>$, and $\mathcal{I}(\bm{\varphi})$ is the Fisher information matrix with its $ij$-th entry given by \cite{Fisher:2008gt,Kay:1993:FSS:151045}
\begin{equation}
  [\mathcal{I}(\bm{\varphi})]_{ij}=\sum_{x\in\Omega}\frac{\partial \ln p_{\bm{\varphi}}(x)}{\partial \varphi_i} \frac{\partial \ln p_{\bm{\varphi}}(x)}{\partial \varphi_j} p_{\bm{\varphi}}(x).
\end{equation}


%
The Fisher information matrix can be further bounded by the quantum Fisher information matrix (QFIM), which leads to the quantum Cram\'er-Rao bound (QCRB)
\cite{helstrom1976quantum,Holevo:2011wv,Braunstein:1994jl} as
\begin{equation}
  Cov(\bm{\hat{\varphi}})\ge\mathcal{I}(\bm{\varphi})^{-1}\ge J(\bm{\varphi})^{-1},
\end{equation}
here the $ij$-th entry of the QFIM is given by
\begin{equation}
  [J(\bm{\varphi})]_{ij}=\frac{1}{2}Tr(\rho_{\bm{\varphi}}\{L_i,L_j\}),
\end{equation}
where $\{\bullet,\bullet\}$ is the anticommutator, $L_i$ is the symmetrical logarithm derivative (SLD) of the $i$-th parameter, which can be obtained by $\partial_{\varphi_i}\rho_{\bm{\varphi}}=(L_i \rho_{\bm{\varphi}}+\rho_{\bm{\varphi}}L_i)/2$. For the multi-parameter quantum estimation, the quantum Cramer-Rao bound is in general not achievable. However, if it satisfies the weaker commutation condition as
\begin{equation}\label{eq:weakercondition}
  Tr(\rho_{\bm{\varphi}}[L_i,L_j])=0,\quad \forall i,j,
\end{equation}
then the quantum Cramer-Rao bound can be saturated\cite{Matsumoto:2002ev,Ragy2016,Pezze:2017ez,2018arXiv180607337Y}.

One distinct feature of the multi-parameter quantum estimation is the trade-off among the precisions of different parameters, as the optimal probe state and the optimal measurement for different parameters are typically different. In general this is treated by adding a weight matrix, $W$, in the figure of merit as $Tr(WCov(\bm{\hat{\varphi}}))$, then one can identify the optimal probe state and the optimal measurement according to the figure of merit. In this article, we will take $W=I$ and consider the figure of merit as $Tr(Cov(\bm{\hat{\varphi}}))$. From the QCRB, it is easy to get a lower bound as
\begin{equation}\label{eq:QCRBmerit}
Tr(Cov(\bm{\hat{\varphi}}))\ge Tr( J(\bm{\varphi})^{-1}).
\end{equation}

\section{Multi-parameter estimation in a three-level system}\label{sec:QCRB}

%

\begin{figure}
\centering
  \subfigure
  {
  \includegraphics[width=0.15\textwidth]{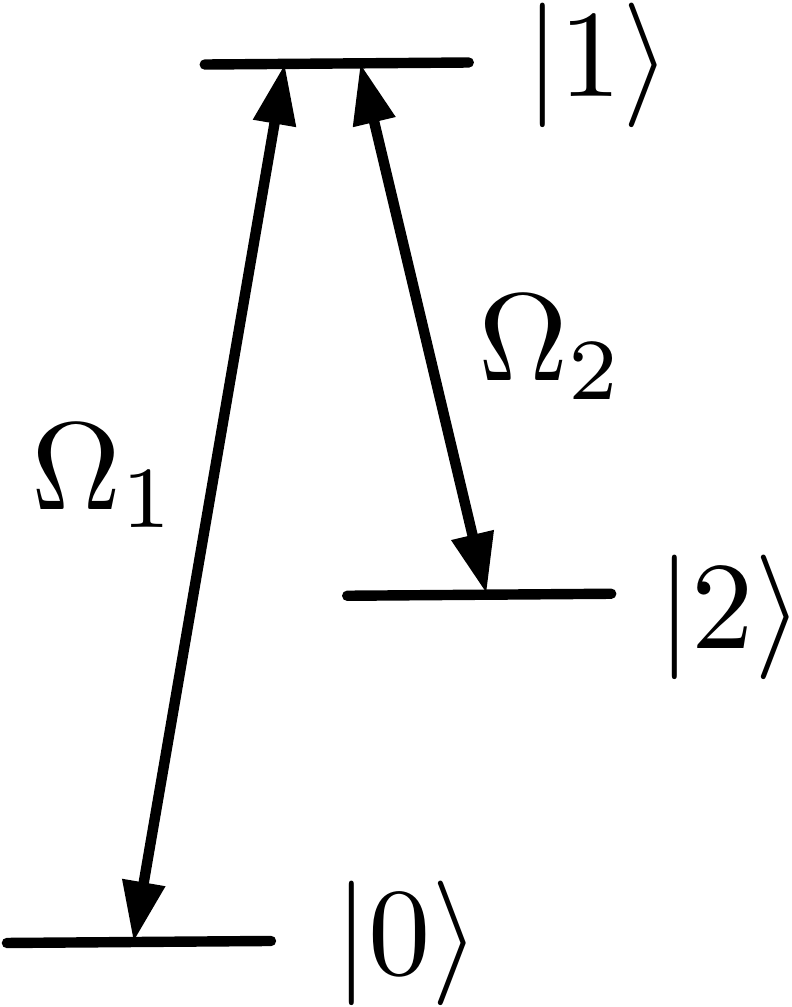}
  }
  \qquad
  \subfigure
  {
  \includegraphics[width=0.09\textwidth]{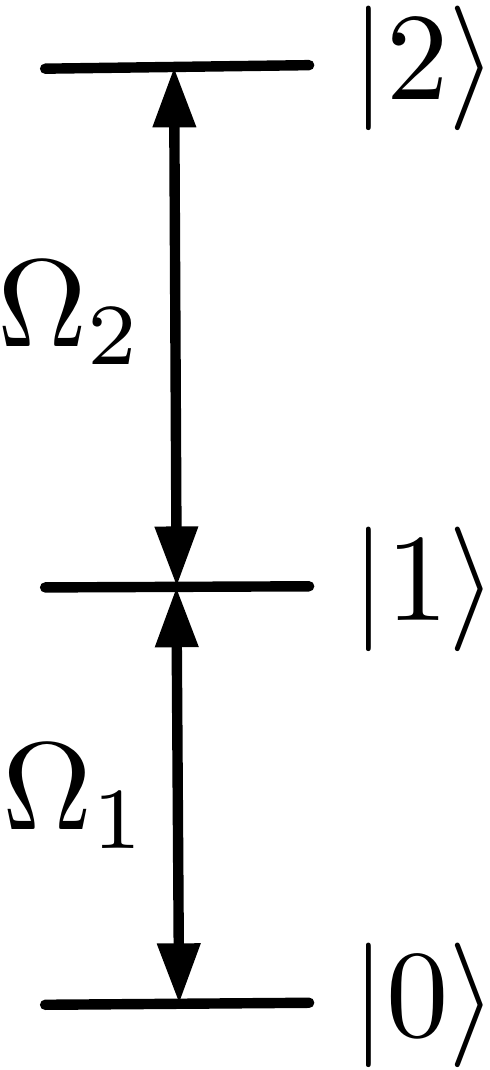}
  }
\caption{Three level systems with $\Lambda$ or ladder shape linkage}
\label{fig:energylevel}
\end{figure}

We first consider a three-level system, with $\Lambda$ or ladder shape linkage, interacting with two resonating radiation fields as depicted in FIG. \ref{fig:energylevel}.

The dynamics of the system is described by the Schr\"{o}dinger equation
\begin{equation}\label{eq:Schordingereq}
  i\hbar\frac{d}{dt}\ket{\psi}=H\ket{\psi},
\end{equation}
where $H$ is the Hamiltonian for the system and its interaction with the radiation fields.
Under the rotating-wave approximation (RWA), the Hamiltonian can be written as
\begin{equation}\label{eq:H}
  H(\Omega)=\frac{1}{2}\begin{pmatrix}
  0 & \Omega_1 & 0\\
  \Omega_1 & 0 & \Omega_2\\
  0 & \Omega_2 & 0
  \end{pmatrix},
\end{equation}
here $\Omega_1$ and $\Omega_2$ are the amplitudes of the fields which are the parameters we would like to estimate.



If we take the initial state as $\ket{\psi_{in}}$, then the output state is given by $\ket{\psi_\Omega}=U(\Omega,t)\ket{\psi_{in}}$, where $U(\Omega,t)=e^{-\mathrm{i}H(\Omega)t}$.
The SLD operators with respect to $\Omega_1$ and $\Omega_2$ can be computed as
\begin{equation}\label{eq:pureSLD}
  L_i=2(\ket{\partial_i\psi_\Omega}\bra{\psi_\Omega}+\ket{\psi_\Omega}\bra{\partial_i\psi_\Omega}),
\end{equation}
where $\partial_i$ denotes the partial derivative with respect to the $i$-th parameter $\Omega_i$,
  $\ket{\partial_i\psi_\Omega}=\frac{\partial U(\Omega,t)}{\partial \Omega_i}\ket{\psi_{in}}.$
The quantum Fisher information matrix is then given by
\begin{equation}\label{eq:FIMelement}
  \begin{aligned}
    \left[J_{\Omega}\right]_{ij}=&\frac{1}{2}\mathrm{Tr}(\ket{\psi_\Omega}\bra{\psi_\Omega}\{L_i,L_j\})\\
    =&2(\inp{\partial_i\psi_\Omega}{\partial_j\psi_\Omega}+\inp{\partial_j\psi_\Omega}{\partial_i\psi_\Omega})\\
    &+4\inp{\partial_i\psi_\Omega}{\psi_\Omega}\inp{\partial_j\psi_\Omega}{\psi_\Omega},
  \end{aligned}
\end{equation}
where $[J_{\Omega}]_{ij}$ is the $ij$-th entry of the quantum Fisher information matrix.
For pure states, the weaker conditions Eq.(\ref{eq:weakercondition}) for saturating the QCRB is reduced to $\mathrm{Im}\inp{\partial_i\psi_\Omega}{\partial_j\psi_\Omega}=0$, $\forall i,j$ \cite{Matsumoto:2002ev,Pezze:2017ez}.


To optimize the figure of merit, one needs to identify the optimal probe state and the optimal measurement. We will first identify the optimal probe state that has the minimal $Tr(J(\bm{\varphi})^{-1})$, then show the quantum Cramer-Rao bound given in Eq.(\ref{eq:QCRBmerit}) can be saturated, ensuring the optimal figure of merit.

To ease the calculation, we will use the eigenvectors of the Hamiltonian as a basis. These eigenbasis $\ket{\Phi_0}$, $\ket{\Phi_{\pm}}$ and the corresponding eigenvalues $\Omega_0$, $\Omega_{\pm}$ of $H(\Omega)$ are as following:
\begin{equation}
\label{eq:eigen}
  \begin{aligned}
    \Omega_0&: \ket{\Phi_0}=\cos\vartheta\ket{0}-\sin\vartheta\ket{2}, \\
    \Omega_+&: \ket{\Phi_+}=\frac{1}{\sqrt{2}}(\sin\vartheta\ket{0}+\ket{1}+\cos\vartheta\ket{2}),\\
    \Omega_-&: \ket{\Phi_-}=\frac{1}{\sqrt{2}}(\sin\vartheta\ket{0}-\ket{1}+\cos\vartheta\ket{2}),
  \end{aligned}
\end{equation}
where $\Omega_0=0$, $\Omega_{\pm}=\pm\frac{1}{2}\sqrt{\Omega_1^2+\Omega_2^2}$, $\vartheta$ is the mixing angle with $\tan\vartheta=\Omega_1/\Omega_2$.

The initial probe state can be written as $\ket{\psi_{in}}=C_0\ket{\Phi_0}+C_+\ket{\Phi_+}+C_-\ket{\Phi_-}$, where $C_0\in\mathbb{R}$, $C_\pm\in\mathbb{C}$ are coefficients to be optimized.
Note that this is just a way to expand the initial probe state in the eigenbasis, and the initial probe state is fixed once it is chosen.
The output state can now be easily obtained as
\begin{equation}\label{eq:output}
  \begin{aligned}
    \ket{\psi_\Omega}&=e^{-\mathrm{i}H(\Omega)t}\ket{\psi_{in}}=\left(\sum_{n=0,\pm}e^{-\mathrm{i}\Omega_n t}\ket{\Phi_n}\bra{\Phi_n}\right)\ket{\psi_{in}}\\
    &=\sum_{n=0,\pm}C_n e^{-\mathrm{i}\Omega_n t}\ket{\Phi_n}.
  \end{aligned}
\end{equation}

The partial derivative of the output state can then be computed as
\begin{equation}\label{eq:psiomega}
  \begin{aligned}
    \ket{\partial_i\psi_\Omega}=&\left(\sum_{n=0,\pm}\partial(e^{-\mathrm{i}\Omega_n t}\ket{\Phi_n}\bra{\Phi_n})/\partial \Omega_i\right)\ket{\psi_{in}}\\
    =&\sum_{n=0,\pm} [(\inp{\partial_i\Phi_n}{\psi_{in}}-\mathrm{i}tC_n\partial_i\Omega_n) e^{-\mathrm{i}\Omega_n t}\ket{\Phi_n}\\
    &+C_n e^{-\mathrm{i}\Omega_n t}\ket{\partial_i\Phi_n}].
  \end{aligned}
\end{equation}
This can then be used to get the quantum Fisher information matrix via Eq.(\ref{eq:FIMelement}) (see the appendix for detailed calculations). The quantum Fisher information matrix turns out to have different behaviours at different time points. There exist some specific time points at which the quantum Fisher information matrices are singular while at other time points the quantum Fisher information matrices are full rank.

Specifically at the time points where $\Omega_+t=2n\pi$ for $n\in\mathbb{N}$, the quantum Fisher information matrix is singular, which takes the form
\begin{equation}
  \begin{aligned}
    J_{\Omega}=&4t^2(1-C_0^2-(|C_+|^2-|C_-|^2)^2)\times\\
    &\begin{pmatrix}
       (\partial_{1} \Omega_+ )^2 &  (\partial_{1} \Omega_+ ) (\partial_{2} \Omega_+ )\\
       (\partial_{2} \Omega_+ ) (\partial_{1} \Omega_+ ) &  (\partial_{2} \Omega_+ )^2
    \end{pmatrix},
  \end{aligned}
\end{equation}
Intuitively, at these time points $U(\Omega, t)=e^{-iH(\Omega) t}=I$, the dynamics is always the Identity operator as long as $\Omega_+$, which corresponds to the norm of $(\Omega_1,\Omega_2)$, takes the specific value, i.e., one can not tell the differences between different pairs of $\Omega_1$ and $\Omega_2$ as long as they have the same specific norm. This indicates that at these time points joint estimations of $\Omega_1$ and $\Omega_2$ are not possible.

At other time points, the entries of the quantum Fisher information matrix are given by(see the appendix for detailed calculations)
\begin{widetext}
  \begin{equation}\label{eq:FIMelementexplicit}
    \begin{aligned}
      \left[J_\Omega\right]_{ij}
      =&2(\inp{\partial_i\psi_\Omega}{\partial_j\psi_\Omega}+\inp{\partial_j\psi_\Omega}{\partial_i\psi_\Omega})+4\inp{\partial_i\psi_\Omega}{\psi_\Omega}\inp{\partial_j\psi_\Omega}{\psi_\Omega}\\
      = &  (\partial_{i} \vartheta ) (\partial_{j} \vartheta )A+ (\partial_{i} \Omega_+ ) (\partial_{j} \Omega_+ )B+\left( (\partial_{i} \vartheta ) (\partial_{j} \Omega_+ )+ (\partial_{j} \vartheta ) (\partial_{i} \Omega_+ )\right)C,
    \end{aligned}
  \end{equation}
\end{widetext}
here $A=-8C_0^2 \mathrm{Im}^2 M +4C_0^2|P|^2+2|M|^2$, $B=-4t^2\frac{\mathrm{Re}^2(MN^*)}{|P|^4}+4t^2-4C_0^2 t^2$, $C=-4\sqrt{2}C_0 \mathrm{Im} M \frac{\mathrm{Re}(MN^*)}{|P|^2}t+2\sqrt{2}C_0 t \mathrm{Im} N$, with $P=e^{-\mathrm{i}\Omega_+t}-1$, $M=C_+^* P^* + C_-^* P$, $N=C_+^* P^* - C_-^* P$.
With this it is then straightforward to get
\begin{widetext}
  \begin{equation}\label{eq:traceofinverse}
    \begin{aligned}
      \mathrm{Tr}(J_\Omega^{-1})
      =&\frac{( (\partial_{1} \vartheta ) (\partial_{1} \vartheta )+ (\partial_{2} \vartheta ) (\partial_{2} \vartheta ))A+2( (\partial_{1} \vartheta ) (\partial_{1} \Omega_+ )+ (\partial_{2} \vartheta ) (\partial_{2} \Omega_+ ))C+( (\partial_{1} \Omega_+ ) (\partial_{1} \Omega_+ )+ (\partial_{2} \Omega_+ ) (\partial_{2} \Omega_+ ))B}{( (\partial_{1} \vartheta ) (\partial_{2} \Omega_+ )- (\partial_{2} \vartheta ) (\partial_{1} \Omega_+ ))^2(AB-C^2)}\\
      =&\frac{(1/4\Omega_+^2) A+ (1/4)B}{(1/16\Omega_+^2)(AB-C^2)}=\frac{4A}{AB-C^2}+\frac{4\Omega_+^2 B}{AB-C^2}.
    \end{aligned}
  \end{equation}
\end{widetext}
In the appendix, we show that the minimal value of $\mathrm{Tr}(J_\Omega^{-1})$ is achieved with the initial probe state taken as $\ket{\phi_{in}}=\frac{P^*}{\sqrt{2}|P|}\ket{\Phi_+}+\frac{P}{\sqrt{2}|P|}\ket{\Phi_-}$, with the corresponding minimal $\mathrm{Tr}(J_\Omega^{-1})$ as
\begin{equation}
\min \mathrm{Tr}(J_\Omega^{-1})=\frac{1}{t^2}+\frac{\Omega_+^2}{4\sin^2(\Omega_+t/2)}.
\end{equation}
It can be checked directly that the weaker commutative conditions $\mathrm{Im}\inp{\partial_i\psi_\Omega}{\partial_j\psi_\Omega}=0$, $\forall i,j$, hold in this case, and thus this minimal value can be saturated by some proper measurements. For example, one set of the measurements that saturates the quantum Cramer-Rao bound can be chosen as $\Pi_1=\ket{\gamma_1}\bra{\gamma_1}$, $\Pi_2=\ket{\gamma_2}\bra{\gamma_2}$, $\Pi_3=\ket{\gamma_3}\bra{\gamma_3}$ and $\Pi_4=I-\Pi_1-\Pi_2-\Pi_3$, where
\begin{equation}\label{eq:optmeasurement}
  \begin{aligned}
    \ket{\gamma_1}=&\ket{\psi_\Omega},\\
    \ket{\gamma_2}=&\frac{\ket{\partial_1\psi_\Omega}}{\sqrt{\inp{\partial_1\psi_\Omega}{\partial_1\psi_\Omega}}},\\
    \ket{\gamma_3}=&\frac{\inp{\partial_2\psi_\Omega}{\gamma_2}\ket{\gamma_2}-\ket{\partial_2\psi_\Omega}}{\sqrt{\inp{\partial_2\psi_\Omega}{\partial_2\psi_\Omega}-\inp{\partial_2\psi_\Omega}{\gamma_2}^2}}.
  \end{aligned}
\end{equation}

\section{Comparison between the joint estimation and the separate estimation}\label{sec:comparison}
We compare the performances of the joint estimation with the separate estimation. The separate estimation is to estimate $\Omega_1$ and $\Omega_2$ separately. For each round of experiments, one can use the dynamical decoupling to remove one of the fields, and reduce the problem to the single parameter estimation. For example, by applying periodic $\pi$-pulses along the direction $(\ket{0}\bra{1}+\ket{1}\bra{0})/2$, one can remove $\Omega_2$ in the Hamiltonian and get the effective Hamiltonian as $H=\Omega_1 H_1=\Omega_1(\ket{0}\bra{1}+\ket{1}\bra{0})/2$.
The quantum Fisher information (QFI) for the single parameter, which has been studied extensively \cite{Giovannetti:2006cr,Giovannetti:2011jk}, is given by
\begin{equation}
  J(\rho(t))=4t^2\left(Tr(\rho_{in}H_1^2)-(Tr(\rho_{in}H_1)^2)\right),
\end{equation}
where $\rho_{in}$ is the input state.
The optimal input state that maximizes the quantum Fisher information is \cite{Giovannetti:2006cr,Giovannetti:2011jk}
\begin{equation}
  \ket{\psi_{in}}=\frac{1}{\sqrt{2}}(\ket{\lambda_{\min}}+\ket{\lambda_{\max}}),
\end{equation}
where $\lambda_{\min}$ and $\lambda_{\max}$ are the minimal and maximal eigenvalues of $H_1$ with $\ket{\lambda_{\min}}$ and $\ket{\lambda_{\max}}$ denoting the corresponding eigenstates.
For $H_1=(\ket{0}\bra{1}+\ket{1}\bra{0})/2$, the maximal QFI is $J=t^2$. Similarly, the maximal QFI for the separate estimation of $\Omega_2$ is also $J=t^2$.

For comparison, we assume that the experiments are repeated $m$ times for the estimation of each parameter in the separate estimation. For the joint estimation, the experiments are repeated $2m$ times so the total number of experiments are the same. The quantum Cram\'er-Rao bound for the joint estimation is given by
\begin{equation}
\label{eq:variance}
  \begin{aligned}
    &Tr(Cov(\hat{\Omega}))\ge \frac{1}{2m}Tr(J_{\Omega}^{-1})\\
    &= \frac{1}{2mt^2}+\frac{\Omega_+^2}{8m\sin^2(\Omega_+t/2)}.
  \end{aligned}
\end{equation}
For the separate estimation, the minimal variance given by the quantum Cramer-Rao bound is $Var(\hat{\Omega}_i)\geq \frac{1}{mt^2}$, $i\in \{1,2\}$. The total variance is
\begin{equation}
\label{eq:separate}
  Tr(Cov(\hat{\Omega}))=Var(\hat{\Omega}_1)+Var(\hat{\Omega}_2)\ge \frac{2}{mt^2}.
\end{equation}
\begin{figure}
  \includegraphics[width=0.5\textwidth]{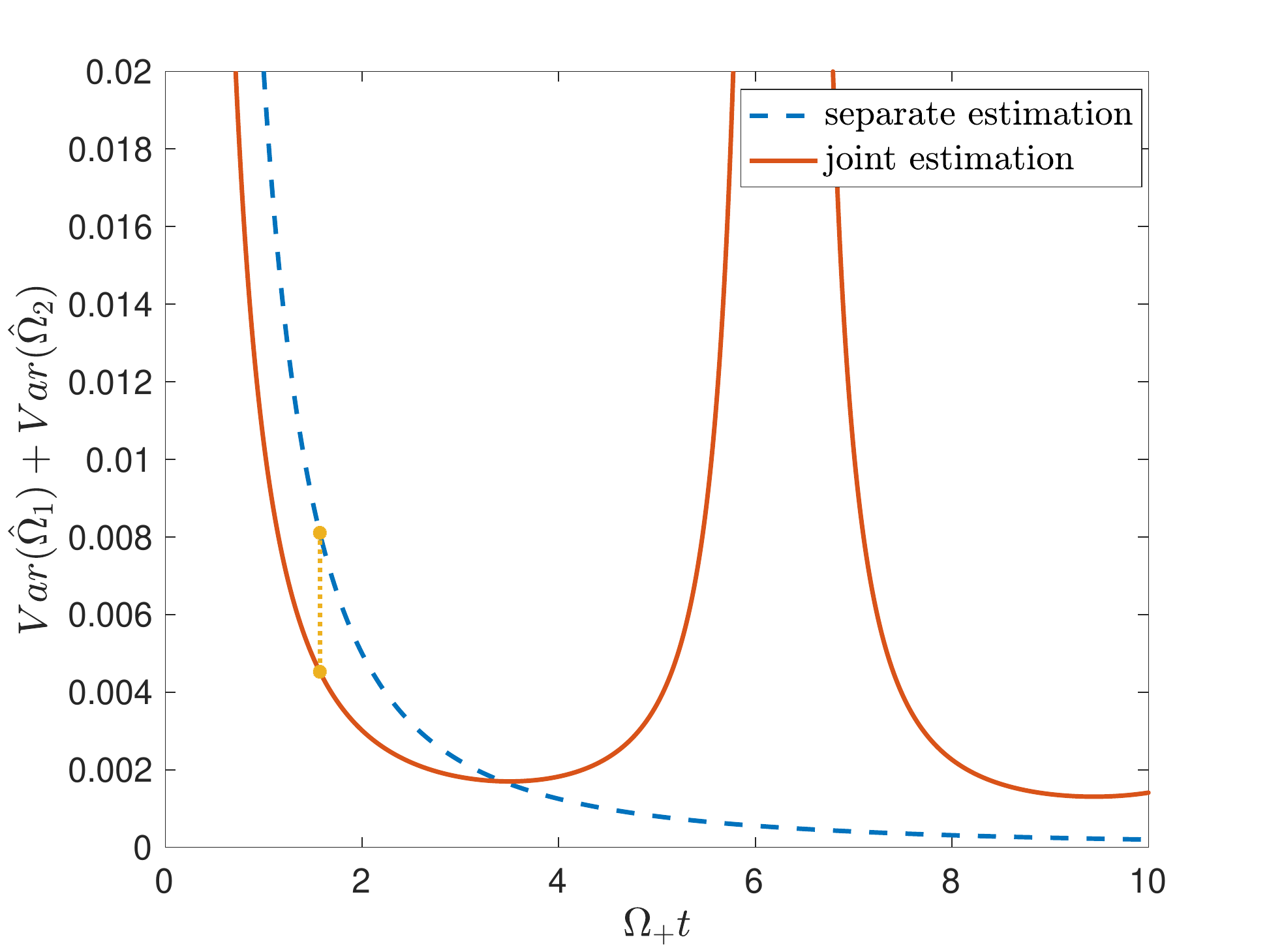}
  \caption{Comparison of joint estimation (solid line) of multiple frequencies with the separate estimation (dashed line), where the $x$-axis is $\Omega_+ t$ and $y$-axis is the total variance $Var(\hat{\Omega}_1)+Var(\hat{\Omega}_2)$. (Without loss of generality, here we set $\Omega_+=0.1$ and $m=1$.) The dotted vertical line shows the difference between these two schemes at the point of $\Omega_+ t=\pi/2$.}
  \label{fig:comparasion}
\end{figure}
The performances of the two schemes can be seen in FIG.\ref{fig:comparasion}, which shows that in the short time regime the joint estimation outperforms the separate estimation while in the long time regime the joint estimation under-performs the separate estimation.

\section{Estimation with the adaptive control}\label{sec:control}

\begin{figure}
  \includegraphics[width=0.45\textwidth]{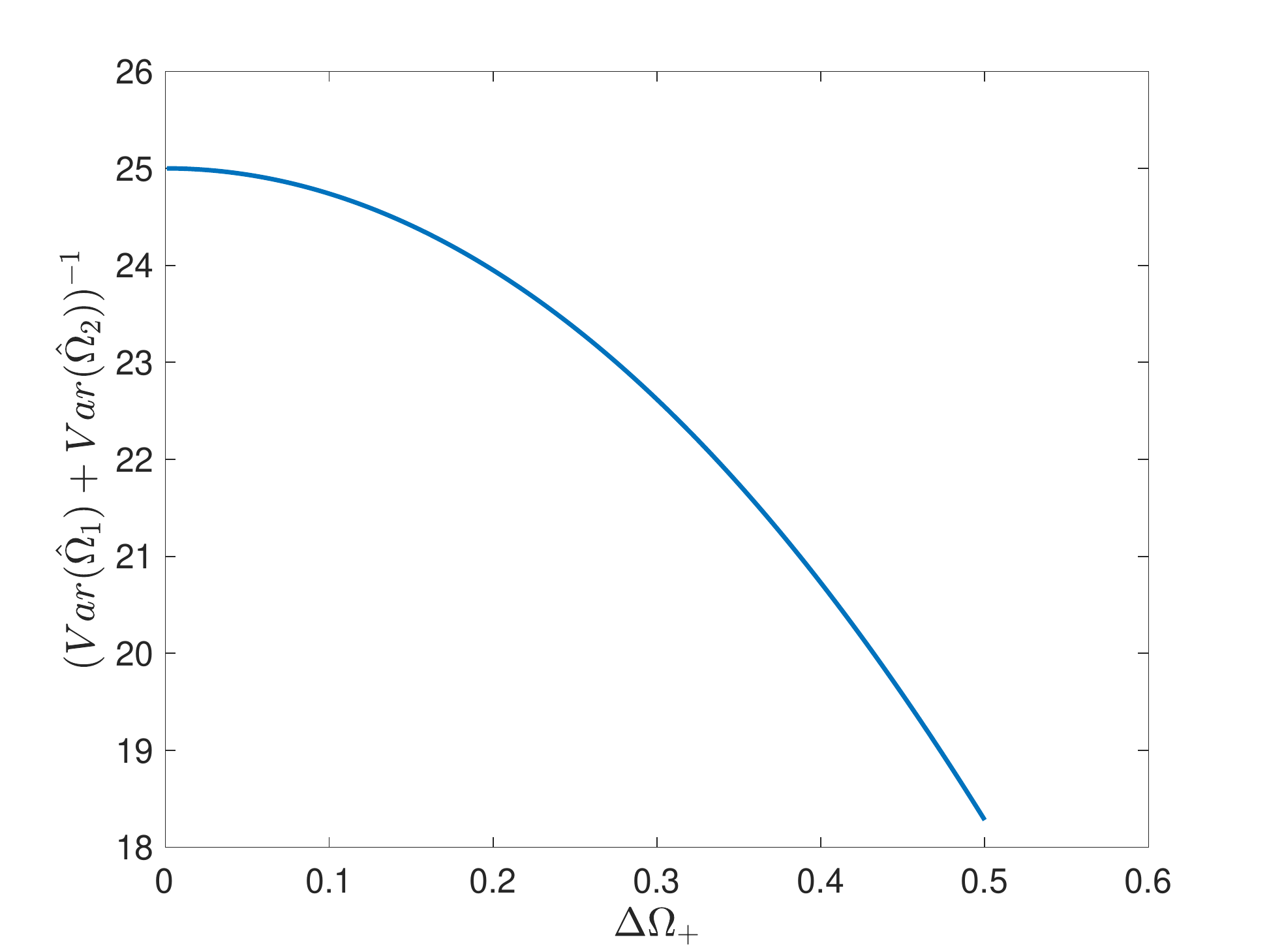}
  \caption{The variation of the total variance at time $t=5$ with respect to the estimation error. Here the $x$-axis is the estimation error, $\Delta\Omega_+\equiv \sqrt{(\Omega_1-\hat{\Omega}_1)^2+(\Omega_2-\hat{\Omega}_2)^2}$, $y$-axis is the inverse of the total variance $(Var(\hat{\Omega}_1)+Var(\hat{\Omega}_2))^{-1}$. It shows that the total variance is quite robust against the estimation error, it changes quite slowly with the estimation error.}
  \label{fig.fig4}
\end{figure}

\begin{figure}
  \includegraphics[width=0.45\textwidth]{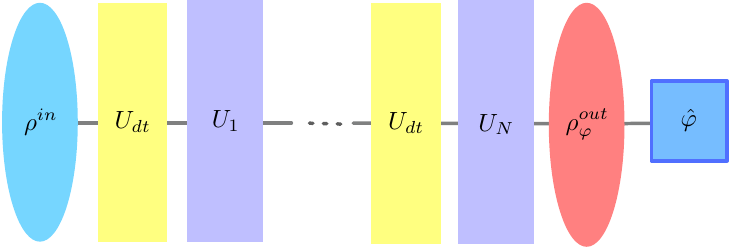}
  \caption{Sequential unitary control}
  \label{fig:sq_control}
\end{figure}

So far we considered the highest precision achievable by optimizing the probe state and the measurement, while assuming the dynamics is fixed as $U(\Omega, t)=e^{-iH(\Omega)t}$. In practice, however, additional controls can often be added during the evolutions as depicted in FIG. \ref{fig:sq_control}, where the total evolution time $t$ has been divided into $N$ intervals with $dt=t/N$ and the adaptive controls $U_1$, $U_2$, ..., $U_N$ are added sequentially.
For such control-enhanced sequential scheme, it has been shown that for any unitary evolution $U(\Omega,t)=e^{-\mathrm{i}H(\Omega)t}$, the optimal controls are given by $U_1=U_2=\cdots=U_N=U^\dagger(\Omega,dt)$, where $\Omega$ is the true value of the parameter\cite{Yuan:2016hf}. In practice, the true value is not known a priori, so the controls can only be taken as $U_1=U_2=\cdots=U_N=U^\dagger(\hat{\Omega},dt)$, with $\hat{\Omega}$ being the estimated value and need to be updated adaptively when more data are collected.
When the estimated value is close to the true value, the total evolution can be described as
\begin{equation}
  \begin{aligned}
    (U_1 U_{dt})^N&=(e^{\mathrm{i}H(\hat{\Omega})dt}e^{-\mathrm{i}H(\Omega)dt})^N\\
    &\approx [(I+\mathrm{i}H(\hat{\Omega})dt)(I-\mathrm{i}H(\Omega)dt)]^N\\
    &\approx (I-\mathrm{i}(H(\Omega-\hat{\Omega}))dt)^N\\
    &\approx e^{-\mathrm{i}H(\Omega-\hat{\Omega})t},
  \end{aligned}
\end{equation}
Therefore, the effective Hamiltonian of the total dynamics is $H(\Omega-\hat{\Omega})$.
While the estimator is being updated, the change of the total variance is depicted in FIG. \ref{fig.fig4}.
It can be seen that the total variance is quite robust against the estimation error.
Asymptotically, when the estimated value converges to the true value, i.e., $\hat{\Omega}=\Omega$, the adaptive control converges to the optimal control and the problem reduces to the joint estimation of $\Omega=(\Omega_1,\Omega_2)$ at the point $(0,0)$. At this point the optimal probe state is $\ket{\psi_{opt}}=\ket{1}$, and the total variance can be obtained by taking the limit $\Omega_+\rightarrow 0$ in Eq.(\ref{eq:variance}) as
\begin{equation}
 Tr(Cov(\hat{\Omega}))=\frac{1}{mt^2},
\end{equation}
which always has a 2-fold improvement over the separate estimation.
By taking the limit $(\Omega_1,\Omega_2)\rightarrow (0,0)$ in Eq.(\ref{eq:optmeasurement}), one can verify that the optimal measurement saturating the QCRB can be chosen as $\mathcal{M}\equiv\{\ket{0}\bra{0},\ket{1}\bra{1},\ket{2}\bra{2}\}$.

When $\Delta\Omega\equiv(\Delta\Omega_1,\Delta\Omega_2)=(\Omega_1-\hat{\Omega}_1,\Omega_2-\hat{\Omega}_2)$ are not zero, the dynamics has the effective Hamiltonian $H(\Delta\Omega)$, the total variance can be directly obtained from Eq.(\ref{eq:variance}) as
\begin{equation}
  \begin{aligned}
    &Tr(Cov(\hat{\Omega}))\\
    =& \frac{1}{2mt^2}+\frac{\Delta\Omega_1^2+\Delta\Omega_2^2}{8m\sin^2(\sqrt{\Delta\Omega_1^2+\Delta\Omega_2^2}t/2)}\\
    =& \frac{1}{mt^2}+\frac{\Delta\Omega_1^2+\Delta\Omega_2^2}{24m}+\mathcal{O}(\Delta\Omega^4).
  \end{aligned}
\end{equation}

To see how quickly $\hat{\Omega}$ approaches its true value under the adaptive control, we perform a numerical simulation of the adaptive estimation as follows:
\begin{enumerate}
  \item Choose an initial estimation, either randomly or based on some prior knowledge, as $(\hat{\Omega}_1^{(0)},\hat{\Omega}_2^{(0)})$, and construct the control, $U^{\dagger}(\hat{\Omega}^{(0)},dt)$.
  \item Choose the initial state as $\ket{1}$, and let the system evolve for a period of time t, perform the projective measurement, $\mathcal{M}=\{\ket{0}\bra{0},\ket{1}\bra{1},\ket{2}\bra{2}\}$, on the state. Repeat the step for $k=30$ times
  \item Based on all the collected measurement results, use the maximum likelihood estimator to update the estimation to $(\hat{\Omega}_1^{(1)},\hat{\Omega}_2^{(1)})$,  then update the control based on the new estimation.
  \item  Repeat Steps 2-3. 
\end{enumerate}

\begin{figure}
  \includegraphics[width=0.5\textwidth]{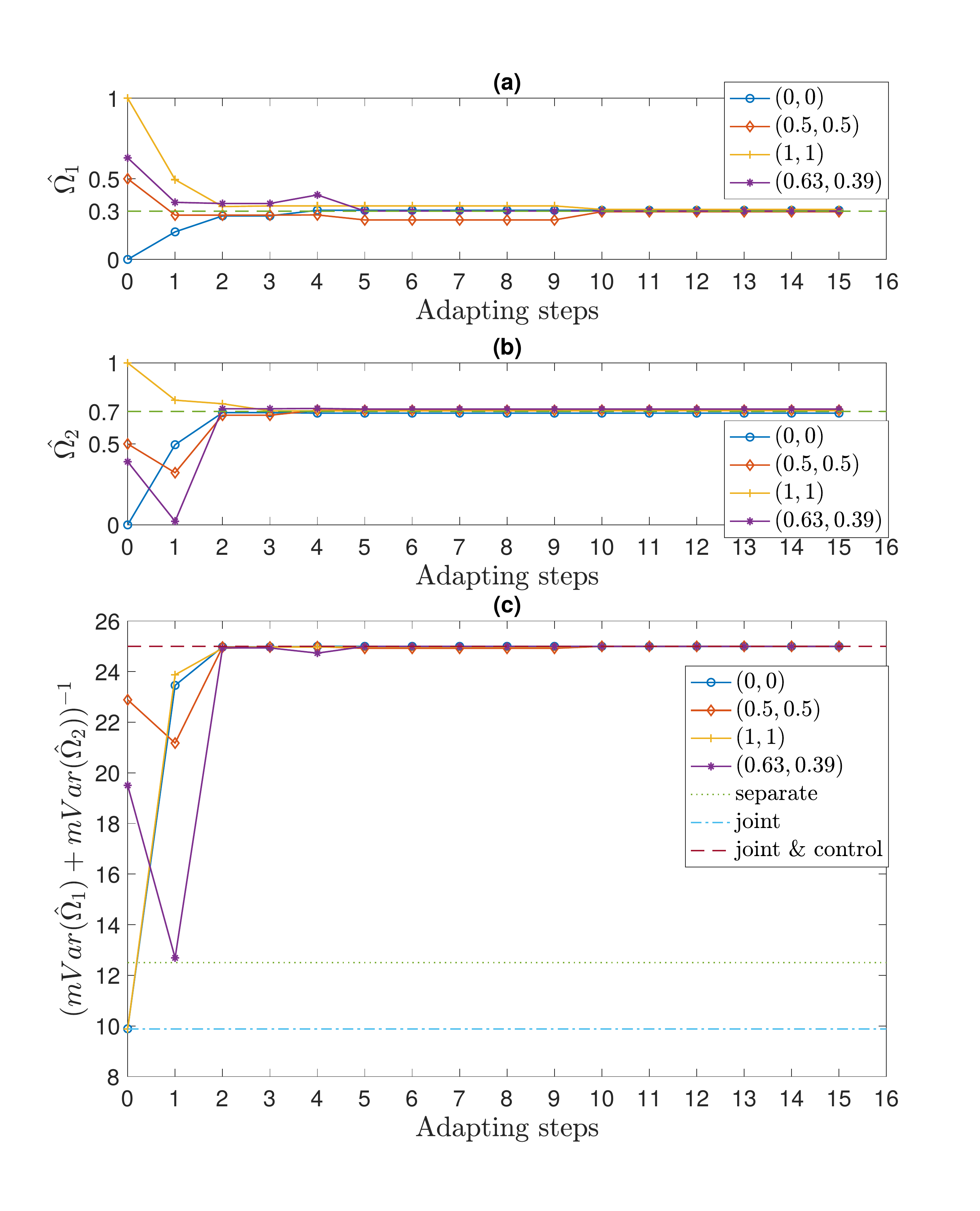}
  \caption{Precision of the estimations at time $t=5$ after different adaptive steps.
  The true values of the parameters are $\Omega_1=0.3$, $\Omega_2=0.7$.
  The $x$-axis is the adapting steps. For subfigures (a) and (b), the $y$-axes are values of the estimators $\hat{\Omega}_1$ and $\hat{\Omega}_2$ respectively.
  The $y$-axis of subfigure (c) is the normalized total variance, i.e. $(mVar(\hat{\Omega}_1)+mVar(\hat{\Omega}_2))^{-1}$ where $m=30n$ with $n$ as the adapting steps.
  The dotted, dash-dotted and dashed lines in (c) represent the total variance for the optimal separate estimation, the optimal joint estimations without control and the optimal joint estimation with control respectively.
  The four solid lines correspond to four different adaptive trajectories with four different initial guesses of the parameters, $(\hat{\Omega}_1,\hat{\Omega}_2)=(0,0), (0.5,0.5),(1,1),(0.63,0.39)$ respectively.
  }
  \label{fig:adap}
\end{figure}

The probe state and the measurement we choose in Step 2 is the optimal probe state and optimal measurement under the optimally controlled scheme. When the estimator approaches the true value, the adaptive scheme converges to the optimal scheme.
It can be seen from FIG. \ref{fig:adap}(a) and (b) that after just a few iterations, the estimators are already close to the true value. FIG. \ref{fig:adap}(c) shows that just after a couple of iteration, the controlled scheme already outperforms the uncontrolled schemes, and after a few iterations the precision under the adaptive control already close to the highest precision achievable under the optimally controlled scheme.

%
\section{Extension to multiple levels}\label{sec:multilevel}
\begin{figure}
\centering
  \includegraphics[width=0.2\textwidth]{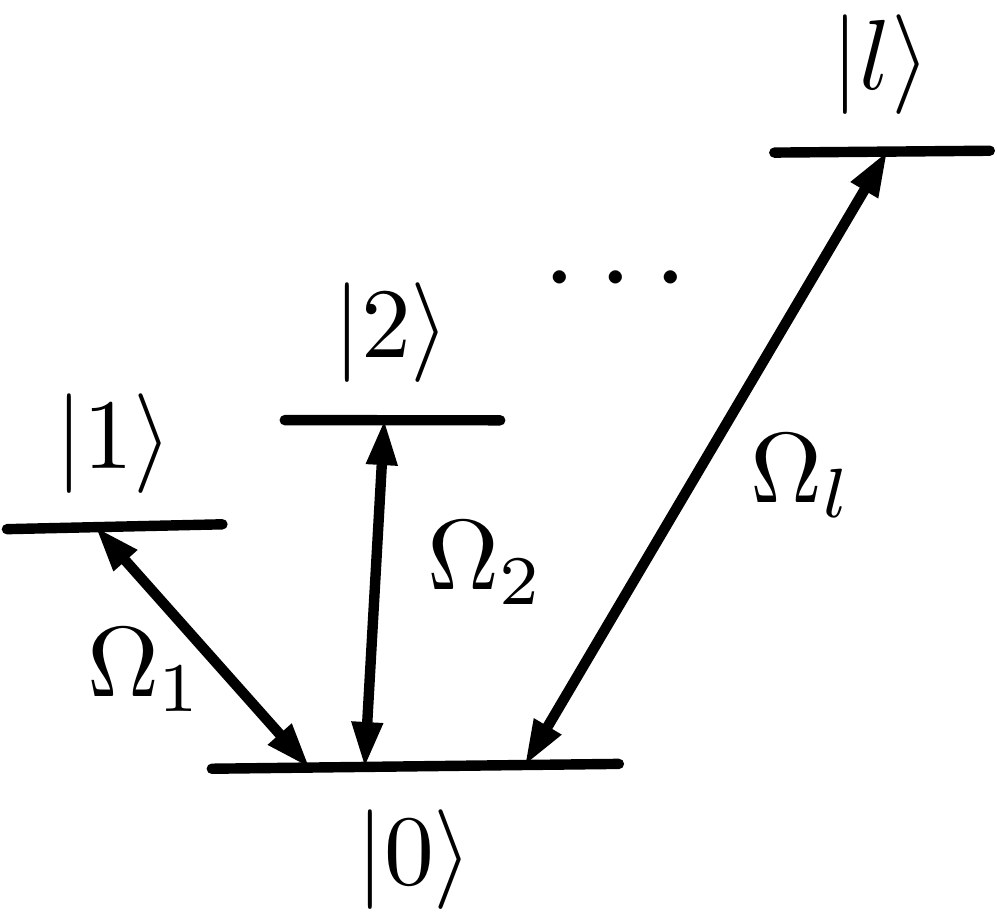}
\caption{$l+1$ energy levels with linkages from one to the other $l$ levels.}
\label{fig.N-level}
\end{figure}

The advantage of the joint estimation with adaptive controls can be extended to systems with more levels.
Considering the system with $l+1$ levels as shown in FIG.\ref{fig.N-level}, we would like to estimate $l$ Rabi frequencies between $\ket{0}$ and the other $l$ energy eigenstates.
In this case the Hamiltonian can be written as $H_l(\Omega)=\sum_{i=1}^l \Omega_i E_{i}$, $E_{i}=(\ket{0}\bra{i}+\ket{i}\bra{0})/2$, where the Rabi frequencies $(\Omega_1,\Omega_2,...,\Omega_l)$ are the parameters to be estimated. For separate estimation, the quantum Fisher information for each parameter is $J=t^2$. If the experiment is repeated $m$ times for each parameter, the total variance becomes
\begin{equation}
\sum_{i=1}^l Var(\hat{\Omega_i})\geq \sum_{i=1}^l \frac{1}{mt^2}=\frac{l}{mt^2}.
\end{equation}

For the joint estimation, by using the optimal adaptive control $U_1=U_2=\cdots=U_N=e^{\mathrm{i}H_l(\hat{\Omega})dt}$ as described in FIG.\ref{fig:sq_control} and choosing the initial state as $\ket{\psi_{in}}=\ket{0}$, we can obtain the quantum Fisher information matrix as $[J_{\Omega}]_{ij}=t^2\delta_{ij}$.
In this case, the weaker commutative condition holds, and thus the quantum Cramer-Rao bound can be saturated. For comparison with the separate estimation, suppose the experiments are repeated for the same total number of times, which is $lm$ times, then the total variance for the joint estimation is
\begin{equation}
\label{eq:joint}
\sum_{i=1}^l Var(\hat{\Omega_i})=Tr(Cov(\hat{\Omega})\geq \frac{1}{lm}Tr(J^{-1})=\frac{1}{mt^2},
\end{equation}
which has an $l$-fold improvement over the separate estimation. Furthermore, if we consider the highest precision achievable for each parameter with the total number of $lm$ experiments, it should be $Var(\hat{\Omega_i})\geq \frac{1}{lmt^2}$. It is generally believed that the highest precision can not be achievable simultaneously for all $l$ parameters as the optimal probe state and the optimal measurement for different parameters are typically incompatible. However, from Eq.(\ref{eq:joint}) we can see that in this case the joint estimation under the adaptive control actually achieves the highest precision for all $l$ parameters simultaneously. No tradeoffs among the precisions of different parameter need to be made. Intuitively, it can be understood that for the estimation of each parameter $\Omega_i$, the state $|0\rangle$ is the optimal state. Without adaptive controls, however, this state is only optimal at the beginning. It will evolve to different states which are no longer optimal. The adaptive controls, on the other hand, can keep the state at the optimal point, thus achieves the highest precision for all parameters simultaneously.


\section{conclusion}\label{sec:conclude}

We have explicitly obtained the optimal probe states and measurements for the joint estimation of multiple Rabi frequencies, and compared with the separate estimation.
These results show that without adaptive controls there exist some time points at which the two Rabi frequencies in a three-level system can not be jointly estimated. There also exist different regimes where joint estimation can either outperform or under-perform separate estimation. However, with additional adaptive controls, the joint estimation can always outperform the separate estimation.
This adds concrete results to the literature on the optimal performance of multi-parameter quantum estimation.

\appendix

\section{Optimal probe state for the joint estimation of two Rabi frequencies}\label{sec:A.state}

 We write the pure input state in the eigenbasis of the Hamiltonian $H$ as
 $\ket{\psi_{in}}=C_0\ket{\Phi_0}+C_+\ket{\Phi_+}+C_-\ket{\Phi_-}$, where $C_0\in\mathbb{R}$, $C_\pm\in\mathbb{C}$ are coefficients to be optimized.
 The evolution of the input state is described by the unitary operator $U=e^{-\mathrm{i}H(\Omega)t}$. Therefore the output state is
 \begin{equation}
   \ket{\psi_\Omega}=e^{-\mathrm{i}H(\Omega)t}\ket{\psi_{in}}=\left(\sum_{n=0,\pm}e^{-\mathrm{i}\Omega_n t}\ket{\Phi_n}\bra{\Phi_n}\right)\ket{\psi_{in}}.
 \end{equation}

 The partial derivative of the state with respect to the $i$-th parameter $\Omega_i$ is given by
 \begin{equation}\label{eq:A.psiomega}
   \begin{aligned}
     \ket{\partial_i\psi_\Omega}=&\left(\sum_{n=0,\pm}\partial(e^{-\mathrm{i}\Omega_n t}\ket{\Phi_n}\bra{\Phi_n})/\partial \Omega_i\right)\ket{\psi_{in}}\\
     =&\sum_{n=0,\pm} (\inp{\partial_i\Phi_n}{\psi_{in}}-\mathrm{i}tC_n\partial_i\Omega_n) e^{-\mathrm{i}\Omega_n t}\ket{\Phi_n}\\
     &+C_n e^{-\mathrm{i}\Omega_n t}\ket{\partial_i\Phi_n},
   \end{aligned}
 \end{equation}
 Note that
 \begin{equation}
   \begin{aligned}
     \ket{\partial_i\Phi_0}=&-\frac{1}{\sqrt{2}} (\partial_{i} \vartheta )(\ket{\Phi_+}+\ket{\Phi_-})\\ \ket{\partial_i\Phi_{\pm}}=&\frac{1}{\sqrt{2}} (\partial_{i} \vartheta )\ket{\Phi_0},
   \end{aligned}
 \end{equation}
 hence
 \begin{equation}\label{eq:A.inpinput}
   \begin{aligned}
     \inp{\partial_i\Phi_0}{\psi_{in}}=&-\frac{1}{\sqrt{2}} (\partial_{i} \vartheta )(C_++C_-)\\ \inp{\partial_i\Phi_{\pm}}{\psi_{in}}=&\frac{1}{\sqrt{2}} (\partial_{i} \vartheta )C_0.
   \end{aligned}
 \end{equation}
 Substituting (\ref{eq:A.inpinput}) back to (\ref{eq:A.psiomega}), we can get the explicit form of $\ket{\partial_i\psi_\Omega}$ expanded by the eigenbasis of $H$:
 \begin{equation}
   \begin{aligned}
     &\ket{\partial_i\psi_\Omega}=\frac{1}{\sqrt{2}} (\partial_{i} \vartheta )\left[C_+(e^{-\mathrm{i}\Omega_+ t}-1)+C_-(e^{-\mathrm{i}\Omega_- t}-1)\right]\ket{\Phi_0}\\
     &+\left[-\mathrm{i}tC_+ (\partial_{i} \Omega_+ ) e^{-\mathrm{i}\Omega_+ t}+\frac{1}{\sqrt{2}}C_0 (\partial_{i} \vartheta )(e^{-\mathrm{i}\Omega_+ t}-1)\right]\ket{\Phi_+}\\
     &+\left[-\mathrm{i}tC_- (\partial_{i} \Omega_- ) e^{-\mathrm{i}\Omega_- t}+\frac{1}{\sqrt{2}}C_0 (\partial_{i} \vartheta )(e^{-\mathrm{i}\Omega_- t}-1)\right]\ket{\Phi_-},
   \end{aligned}
 \end{equation}
 where $i=1,2$, and
 \begin{equation}\label{eq:A.parameterpartial}
   \begin{aligned}
      \partial_{1} \vartheta =\frac{\partial\vartheta}{\partial \Omega_1}=\frac{\Omega_2}{4\Omega_+^2}, \quad  \partial_{2} \vartheta =\frac{\partial\vartheta}{\partial \Omega_2}=-\frac{\Omega_1}{4\Omega_+^2}\\
      \partial_{1} \Omega_+ =\frac{\partial\Omega_+}{\partial \Omega_1}=\frac{\Omega_1}{4\Omega_+},\quad  \partial_{2} \Omega_+ =\frac{\partial\Omega_+}{\partial \Omega_2}=\frac{\Omega_2}{4\Omega_+}.
   \end{aligned}
 \end{equation}


 Recall that the SLD operators for the pure state $\ket{\psi_\Omega}$ can be calculated by $L_i=2(\ket{\partial_i\psi_\Omega}\bra{\psi_\Omega}+\ket{\psi_\Omega}\bra{\partial_i\psi_\Omega})$, $i=1,2$. Thus the entries of SLD Fisher information matrix can be calculated by
 \begin{equation}\label{eq:A.FIMelement}
   \begin{aligned}
     \left[J_\Omega\right]_{ij}=&\frac{1}{2}\mathrm{Tr}(\ket{\psi_\Omega}\bra{\psi_\Omega}\{L_i,L_j\})\\
     =&2(\inp{\partial_i\psi_\Omega}{\partial_j\psi_\Omega}+\inp{\partial_j\psi_\Omega}{\partial_i\psi_\Omega})\\
     &+4\inp{\partial_i\psi_\Omega}{\psi_\Omega}\inp{\partial_j\psi_\Omega}{\psi_\Omega},
   \end{aligned}
 \end{equation}

 Denote $P=e^{-\mathrm{i}\Omega_+t}-1$, $M=C_+^* P^* + C_-^* P$, $N=C_+^* P^* - C_-^* P$. First we consider when $P= 0$. In this case,
 \begin{equation}
   \inp{\partial_i\psi_\Omega}{\psi_\Omega}=\mathrm{i}t (\partial_{i} \Omega_+ )(|C_+|^2-|C_-|^2),
 \end{equation}
 \begin{equation}
   \inp{\partial_i\psi_\Omega}{\partial_j\psi_\Omega}= (\partial_{i} \Omega_+ ) (\partial_{j} \Omega_+ )t^2(1-C_0^2).
 \end{equation}
 Substituting them to (\ref{eq:A.FIMelement}), we can get the QFIM as
 \begin{equation}
   \begin{aligned}
     J_\Omega=&4t^2(1-C_0^2-(|C_+|^2-|C_-|^2)^2)\times\\
     &\begin{pmatrix}
        (\partial_1 \Omega_+ )^2 &  (\partial_1 \Omega_+ ) (\partial_2 \Omega_+ )\\
        (\partial_2 \Omega_+ ) (\partial_1 \Omega_+ ) &  (\partial_2 \Omega_+ )^2
     \end{pmatrix}.
   \end{aligned}
 \end{equation}
 It can be easily seen that the QFIM is singular, indicating that we cannot estimate $\Omega_1$ and $\Omega_2$ simultaneously.

When $P\neq 0$, 
we have
 \begin{equation}\label{eq:A.inp1}
   \begin{aligned}
     \inp{\partial_i\psi_\Omega}{\psi_\Omega}=&\mathrm{i}\sqrt{2} (\partial_{i} \vartheta )C_0 \mathrm{Im}\left[(C_+^* P^* + C_-^* P) \right]\\
     &+\mathrm{i}t (\partial_{i} \Omega_+ )(|C_+|^2-|C_-|^2)\\
     =&\mathrm{i}\sqrt{2} (\partial_{i} \vartheta )C_0 \mathrm{Im} M + \mathrm{i}t (\partial_{i} \Omega_+ ) \frac{\mathrm{Re}(MN^*)}{|P|^2},
   \end{aligned}
 \end{equation}
and
 \begin{widetext}
   \begin{equation}\label{eq:A.inp2}
     \begin{aligned}
       \inp{\partial_i\psi_\Omega}{\partial_j\psi_\Omega}
       =&  (\partial_{i} \Omega_+ ) (\partial_{j} \Omega_+ )t^2 +( (\partial_{i} \vartheta ) (\partial_{j} \vartheta )|P|^2- (\partial_{i} \Omega_+ ) (\partial_{j} \Omega_+ )t^2)C_0^2
       +\frac{1}{2} (\partial_{i} \vartheta ) (\partial_{j} \vartheta )|C_+^* P^* + C_-^* P|^2\\
       &+\mathrm{i}\frac{1}{\sqrt{2}} (\partial_{i} \Omega_+ ) (\partial_{j} \vartheta ) t C_0(C_-^*P-C_+^*P^*)
       +\mathrm{i}\frac{1}{\sqrt{2}} (\partial_{j} \Omega_+ ) (\partial_{i} \vartheta ) t C_0(C_+P-C_-P^*)\\
       =& (\partial_{i} \Omega_+ ) (\partial_{j} \Omega_+ )t^2 +( (\partial_{i} \vartheta ) (\partial_{j} \vartheta )|P|^2- (\partial_{i} \Omega_+ ) (\partial_{j} \Omega_+ )t^2)C_0^2
       +\frac{1}{2} (\partial_{i} \vartheta ) (\partial_{j} \vartheta )|M|^2\\
       &+\mathrm{i}\frac{-1}{\sqrt{2}} (\partial_{i} \Omega_+ ) (\partial_{j} \vartheta ) t C_0 N
       +\mathrm{i}\frac{1}{\sqrt{2}} (\partial_{j} \Omega_+ ) (\partial_{i} \vartheta ) t C_0 N^*.
     \end{aligned}
   \end{equation}
 \end{widetext}
where $\mathrm{Re}x$ and $\mathrm{Im}x$ denote the real and imaginary parts of $x$ respectively. Substituting (\ref{eq:A.inp1}) and (\ref{eq:A.inp2}) back to (\ref{eq:A.FIMelement}), we can get
 \begin{widetext}
   \begin{equation}\label{eq:A.FIMelementexplicit}
     \begin{aligned}
       \left[J_\Omega\right]_{ij}
       =&2(\inp{\partial_i\psi_\Omega}{\partial_j\psi_\Omega}+\inp{\partial_j\psi_\Omega}{\partial_i\psi_\Omega})+4\inp{\partial_i\psi_\Omega}{\psi_\Omega}\inp{\partial_j\psi_\Omega}{\psi_\Omega}\\
       =&  (\partial_{i} \vartheta ) (\partial_{j} \vartheta )\left(-8C_0^2 \mathrm{Im}^2 M +4C_0^2|P|^2+2|M|^2\right)
       +  (\partial_{i} \Omega_+ ) (\partial_{j} \Omega_+ )\left(-4t^2\frac{\mathrm{Re}^2(MN^*)}{|P|^4}+4t^2-4C_0^2 t^2\right)\\
       &+ \left( (\partial_{i} \vartheta ) (\partial_{j} \Omega_+ )+ (\partial_{j} \vartheta ) (\partial_{i} \Omega_+ )\right)
       \left( -4\sqrt{2}C_0 \mathrm{Im} M \frac{\mathrm{Re}(MN^*)}{|P|^2}t+2\sqrt{2}C_0 t \mathrm{Im} N\right)\\
       = &  (\partial_{i} \vartheta ) (\partial_{j} \vartheta )A+ (\partial_{i} \Omega_+ ) (\partial_{j} \Omega_+ )B+\left( (\partial_{i} \vartheta ) (\partial_{j} \Omega_+ )+ (\partial_{j} \vartheta ) (\partial_{i} \Omega_+ )\right)C,
     \end{aligned}
   \end{equation}
 \end{widetext}
 where  $A=-8C_0^2 \mathrm{Im}^2 M +4C_0^2|P|^2+2|M|^2$, $B=-4t^2\frac{\mathrm{Re}^2(MN^*)}{|P|^4}+4t^2-4C_0^2 t^2$, $C=-4\sqrt{2}C_0 \mathrm{Im} M \frac{\mathrm{Re}(MN^*)}{|P|^2}t+2\sqrt{2}C_0 t \mathrm{Im} N$. It then straightforward to get
 \begin{widetext}
   \begin{equation}\label{eq:A.traceofinverse}
     \begin{aligned}
       \mathrm{Tr}(J_\Omega^{-1})
       =&\frac{( (\partial_{P} \vartheta ) (\partial_{P} \vartheta )+ (\partial_{S} \vartheta ) (\partial_{S} \vartheta ))A+2( (\partial_{P} \vartheta ) (\partial_{P} \Omega_+ )+ (\partial_{S} \vartheta ) (\partial_{S} \Omega_+ ))C+( (\partial_{P} \Omega_+ ) (\partial_{P} \Omega_+ )+ (\partial_{S} \Omega_+ ) (\partial_{S} \Omega_+ ))B}{( (\partial_{P} \vartheta ) (\partial_{S} \Omega_+ )- (\partial_{S} \vartheta ) (\partial_{P} \Omega_+ ))^2(AB-C^2)}\\
       =&\frac{(1/4\Omega_+^2) A+ (1/4)B}{(1/16\Omega_+^2)(AB-C^2)}=\frac{4A+4\Omega_+^2 B}{AB-C^2}.
     \end{aligned}
   \end{equation}
 \end{widetext}

Since $|M|^2+|N|^2=2|P|^2(1-C_0^2)$, we can let
 \begin{equation}
   \begin{aligned}
     M=&\sqrt{2|P|^2(1-C_0^2)} e^{\mathrm{i}\alpha}\cos\theta,\\
     N=&\sqrt{2|P|^2(1-C_0^2)} e^{\mathrm{i}\beta}\sin\theta.
   \end{aligned}
 \end{equation}
Then
 \begin{equation}
   \begin{aligned}
     A=&4|P|^2C_0^2+4|P|^2(1-C_0^2)\cos^2\theta\\
     &-16|P|^2C_0^2(1-C_0^2)\cos^2\theta\sin^2\alpha,\\
     B=&4t^2(1-C_0^2)-4t^2(1-C_0^2)^2\sin^2 2\theta\cos^2(\alpha-\beta).
   \end{aligned}
 \end{equation}
With this we are going to show that $\dfrac{A}{AB-C^2}\ge\dfrac{1}{4t^2}$ and $\dfrac{B}{AB-C^2}\ge\dfrac{1}{4|P|^2}$.

To show $\dfrac{A}{AB-C^2}\ge\dfrac{1}{4t^2}$, we note that
 \begin{equation}\label{eq:A.part1}
   \begin{aligned}
     &\frac{A}{AB-C^2}-\frac{1}{4t^2}=\frac{4t^2A-(AB-C^2)}{4t^2(AB-C^2)}\\
     &\ge\frac{A(4t^2-B)}{4t^2(AB-C^2)}.
   \end{aligned}
 \end{equation}
From the fact that the quantum Fisher information matrix is positive semidefinite, we can obtain $AB-C^2\ge 0$.
Also \begin{equation}
   \begin{aligned}
     A\ge &4|P|^2C_0^2\cos^2\theta+4|P|^2(1-C_0^2)\cos^2\theta\\
     &-16|P|^2C_0^2(1-C_0^2)\cos^2\theta\sin^2\alpha\\
     \ge &4|P|^2\cos^2\theta-4|P|^2\cos^2\theta\sin^2\alpha\ge0,
   \end{aligned}
 \end{equation}
 where we used the fact $4C_0^2(1-C_0^2)\le 1$, and
 \begin{equation}
   \begin{aligned}
     4t^2-B= 4t^2C_0^2+4t^2(1-C_0^2)^2\sin^2 2\theta\cos^2(\alpha-\beta)\ge 0.
   \end{aligned}
 \end{equation}
Thus $\frac{A}{AB-C^2}-\frac{1}{4t^2}\geq 0$. The equality can be saturated when $C=0$ and $B=4t^2$.

To show $\dfrac{B}{AB-C^2}\ge\dfrac{1}{4|P|^2}$, we note that
 \begin{equation}\label{eq:A.part2}
   \begin{aligned}
     &\frac{B}{AB-C^2}-\frac{1}{4|P|^2}=\frac{4|P|^2 B-(AB-C^2)}{4|P|^2(AB-C^2)}\\
     &\ge \frac{B(4|P|^2 -A)}{4|P|^2(AB-C^2)}.
   \end{aligned}
 \end{equation}
 Since
 \begin{equation}
   B\ge 4t^2(1-C_0^2)-4t^2(1-C_0^2)^2=4t^2C_0^2(1-C_0^2)\ge 0,
 \end{equation}
 and
 \begin{equation}
   \begin{aligned}
     &4|P|^2 -A\\
     = &4|P|^2(1-C_0^2)(1-\cos^2\theta)\\
     &+16|P|^2C_0^2(1-C_0^2)\cos^2\theta\sin^2\alpha\\
     \ge & 0.
   \end{aligned}
 \end{equation}
 Thus $\frac{B}{AB-C^2}-\frac{1}{4|P|^2}\geq 0$ and the equality can be saturated when $C=0$ and $A=4|P|^2$.

With $\frac{A}{AB-C^2}\ge\dfrac{1}{4t^2}$ and $\frac{B}{AB-C^2}\geq \frac{1}{4|P|^2}$, it is then easy to see that
 \begin{equation}
   \begin{aligned}
     \mathrm{Tr}(J_\Omega^{-1})=&\frac{4A+4\Omega_+^2 B}{AB-C^2}\\
     \ge & \frac{1}{t^2}+\frac{\Omega_+^2}{|P|^2}\\
     =&\frac{1}{t^2}+\frac{\Omega_+^2}{4\sin^2(\Omega_+t/2)},
   \end{aligned}
 \end{equation}
where the equality can be achieved with $A=4|P|^2$, $B=4t^2$ and $C=0$. It is straightforward to check that when the input state takes the form $\ket{\psi_{in}}=P^*/(\sqrt{2}|P|)\ket{\Phi_+}+P/(\sqrt{2}|P|)\ket{\Phi_-}$, the equality is saturated. And thus it is the optimal probe state.
Since $P=e^{-\mathrm{i}\Omega_+t}-1=-2\mathrm{i}\sin(\Omega_+t/2)e^{-\mathrm{i}\Omega_+t/2}$, the optimal input state can also be written as
 \begin{equation}
   \begin{aligned}
     \ket{\psi_{in}}=&\frac{P^*}{\sqrt{2}|P|}\ket{\Phi_+}+\frac{P}{\sqrt{2}|P|}\ket{\Phi_-}\\
     =&\frac{e^{\mathrm{i}\Omega_+t/2}\ket{\Phi_+}-e^{\mathrm{i}\Omega_-t/2}\ket{\Phi_-}}{\sqrt{2}}
   \end{aligned}
 \end{equation}

\end{document}